\documentclass[prl,reprint,showpacs,amsmath,amssymb,superscriptaddress,longbibliography]{revtex4-1}

\usepackage{graphicx,bm,color,mathptmx,hyperref} 

\newcommand{\Tr}{\mathop{\mathrm{Tr}} \nolimits}

\newcommand{\op}[1]{\hat{#1}}

\begin{document}

\title{Extremal quantum states and their Majorana constellations}

\author{G.~Bj\"{o}rk} 
\affiliation{Department of Applied Physics,
  Royal Institute of Technology (KTH), AlbaNova,
  SE-106 91 Stockholm, Sweden}

\author{A.~B.~Klimov} 
\affiliation{Departamento de F\'{\i}sica,
  Universidad de Guadalajara, 44420~Guadalajara, Jalisco, Mexico}

\author{P.~de la Hoz} 
\affiliation{Departamento de \'Optica, Facultad de F\'{\i}sica, 
Universidad Complutense, 28040~Madrid, Spain}

\author{M.~Grassl}
\affiliation{Max-Planck-Institut f\"ur die Physik des Lichts, 
G\"{u}nther-Scharowsky-Stra{\ss}e 1, Bau 24, 91058 Erlangen, Germany} 
\affiliation{ Institut f\"ur Optik, Information und Photonik,
  Universit\"{a}t Erlangen-N\"{u}rnberg, Staudtstra{\ss}e 7/B2,
  91058 Erlangen, Germany}

\author{G.~Leuchs} 
\affiliation{Max-Planck-Institut f\"ur die Physik des Lichts, 
G\"{u}nther-Scharowsky-Stra{\ss}e 1, Bau 24, 91058 Erlangen, Germany} 
\affiliation{ Institut f\"ur Optik, Information und Photonik,
  Universit\"{a}t Erlangen-N\"{u}rnberg, Staudtstra{\ss}e 7/B2,
  91058 Erlangen, Germany}

\author{L.~L.~S\'{a}nchez-Soto} 
\affiliation{Departamento de \'Optica, Facultad de F\'{\i}sica,
 Universidad Complutense, 28040~Madrid,  Spain} 
\affiliation{Max-Planck-Institut f\"ur die Physik des Lichts,
  G\"{u}nther-Scharowsky-Stra{\ss}e 1, Bau 24, 91058 Erlangen,
  Germany} 
\affiliation{ Institut f\"ur Optik, Information und Photonik,
  Universit\"{a}t Erlangen-N\"{u}rnberg, Staudtstra{\ss}e 7/B2,
  91058 Erlangen, Germany}

\begin{abstract}
  The characterization of quantum polarization of light requires
  knowledge of all the moments of the Stokes variables, which are
  appropriately encoded in the multipole expansion of the density
  matrix. We look into the cumulative distribution of those multipoles
  and work out the corresponding extremal pure states. We find that
  SU(2) coherent states are maximal to any order whereas the converse
  case of minimal states (which can be seen as the most quantum ones)
  is investigated for a diverse range of the number of photons.  Taking
  advantage of the Majorana representation, we recast the problem as
  that of distributing a number of points uniformly over the surface
  of the Poincar\'e sphere.
\end{abstract}


\maketitle

\emph{Introduction.---} 
Stokes variables constitute an invaluable tool for assessing light
polarization, both in the classical and quantum domains. They can be
efficiently measured and lead to an elegant geometric representation,
the Poincar\'e sphere, which not only provides remarkable insights,
but also greatly simplifies otherwise complex 
problems.

Classical polarization is chiefly built on first-order moments of the
Stokes parameters: states are pictured as points on the Poincar\'e
sphere (i.e., neglecting fluctuations altogether). Nowadays, however,
there is a general agreement that a thorough understanding of the
effects arising in the realm of the quantum world calls for an
analysis of higher-order polarization
fluctuations~\cite{Klimov:2005kl,Sehat:2005wd,Marquardt:2007bh,
  Klimov:2010uq,Muller:2012ys,Bjork:2012zr,Singh:2013ly}.  In fact,
this is what comes up in coherence theory, where, in general, one
needs a hierarchy of correlation functions to specify a field.
 
Recently, we have laid the foundations for a systematic solution to
this fundamental and longstanding
question~\cite{Sanchez-Soto:2013cr,Hoz:2013om,Hoz:2014kq}.  The
backbone of our proposal is a multipole expansion of the density
matrix, which naturally sorts successive moments of the Stokes
variables. The dipole term, being just the first-order moment, renders
the classical picture, while the other multipoles account for the
fluctuations we wish to scrutinize.  Consequently, the cumulative
distribution for these multipoles yields complete information about
the polarization properties.

This Communication represents a substantial step ahead in this program, as we
elaborate on the extremal states for the aforementioned multipole
distribution. We find that the SU(2) coherent states maximize it to
any order, so they are the most polarized allowed by quantum theory.
We determine as well the states that kill the cumulative distribution
up to a given order $M$: they serve precisely as the opposite of SU(2)
coherent states and hence can be considered as the kings of
quantumness. Furthermore, employing the striking advantages of the
Majorana representation~\cite{Majorana:1932ul}, these kings appear
naturally related to the problem of distributing $N$ points on the
Poincar\'e sphere in the ``most symmetric'' fashion, a problem with a
long history and many different solutions depending on the cost
function one tries to optimize~\cite{Conway:1996ys,Saff:1997gj}.

\emph{Polarization multipoles.---}
The quantum Stokes operators are defined as ~\cite{Luis:2000ys}  
\begin{equation}
  \label{eq:Stokop}
  \begin{array}{c}
    \op{S}_{x} = \textstyle\frac{1}{2} 
    ( \op{a}^{\dagger}_{+}  \op{a}_{-} + 
    \op{a}^{\dagger}_{-} \op{a}_{+} ) \, ,  
    \qquad
    \op{S}_{y} =  \textstyle\frac{i}{2} 
  ( \op{a}_{+} \op{a}^{\dagger}_{-} - 
    \op{a}^{\dagger}_{+} \op{a}_{-} ) \, ,  \\
    \\
    \op{S}_{z}  = \textstyle\frac{1}{2}  
  ( \op{a}^{\dagger}_{+} \op{a}_{+} - 
    \op{a}^{\dagger}_{-} \op{a}_{-} ) \, ,
  \end{array}
\end{equation}
together with the total photon number
$ \op{N} = \op{a}^{\dagger}_{+} \op{a}_{+} + \op{a}^{\dagger}_{-}
\op{a}_{-}$. Here, $\op{a}_{+}$ and $\op{a}_{-}$ represent the 
amplitudes in two circularly-polarized orthogonal modes. We 
have that $[\op{a}_{k}, \op{a}_{\ell}^{\dagger} ] = \delta_{k \ell}$,
$k, \ell \in \{+, - \}$, with $\hbar =1$ throughout and the
superscript $\dagger$ stands for the Hermitian conjugate. 
 The definition (\ref{eq:Stokop}) differs by a factor 1/2 from 
its classical counterpart~\cite{Born:1999yq}, but in this way 
the components of the vector $\op{\mathbf{S}} = 
(\op{S}_{x}, \op{S}_{y}, \op{S}_{z})$ satisfy the su(2) commutation 
relations: $ [ \op{S}_{x}, \op{S}_{y}] = i \op{S}_{z}$ and cyclic
permutations. For an $N$-photon state, $\op{\mathbf{S}}^{2} = 
S (S+1) \op{\openone} $, where $S= N/2$, so the number of 
photons fixes the effective spin.

Put in a different way, (\ref{eq:Stokop}) is nothing but the Schwinger
representation of the su(2) algebra. Consequently, the ideas to be
explored here are by no means restricted to polarization, but concern
numerous instances wherein su(2) is the fundamental
symmetry~\cite{Chaturvedi:2006vn}.
  
In our case, $ [ \op{N}, \op{\mathbf{S}} ] = 0$, so each subspace with
a fixed number of photons ought to be addressed separately.  To bring
out this fact more prominently, instead of the Fock states
$\{ |n_{+}, n_{-} \rangle \} $, we employ the relabeling
$ | S, m \rangle \equiv | n_{+} = S + m, n_{-} = S - m \rangle$, which can
be thought of as the common eigenstates of $ \op{\mathbf{S}}^{2}$ and
$\op{S}_{z}$.  For each fixed $S$, $m$ runs from $-S$ to $S$, 
and the states $\{ |S, m \rangle \} $ span a $(2S+1)$-dimensional
invariant subspace~\cite{Varshalovich:1988ct}.

As a result, the only accessible information from any density matrix
$\op{\varrho}$ is its block-diagonal form
$ \op{\varrho}_{\mathrm{pol}} = \bigoplus_{S} \op{\varrho}^{(S)}$, where
$\op{\varrho}^{(S)}$ is the density matrix in the subspace of spin
$S$. This $\op{\varrho}_{\mathrm{pol}}$ has been termed the
polarization sector~\cite{Raymer:2000zt} or the polarization density
matrix~\cite{Karassiov:2004xw}. It is advantageous  to expand each  
$\op{\varrho}^{(S)}$  as 
\begin{equation}
  \label{rho1}
  \op{\varrho}^{(S)} =  \sum_{K= 0}^{2S} \sum_{q=-K}^{K}  
  \varrho_{Kq}^{(S)} \,   \op{T}_{Kq}^{(S)} \, ,
\end{equation}
rather than using directly the basis $\{ | S, m \rangle \}$.  The
irreducible tensor operators $\op{T}_{Kq}^{(S)}$
are~\cite{Fano:1959ly,Blum:1981rb}
\begin{equation}
  \label{Tensor} 
  \op{T}_{Kq}^{(S)} = \sqrt{\frac{2 K +1}{2 S +1}} 
  \sum_{m,  m^{\prime}= -S}^{S} C_{Sm, Kq}^{Sm^{\prime}} \, 
  |  S , m^\prime \rangle \langle S, m | \, ,
\end{equation}
with $ C_{Sm, Kq}^{Sm^{\prime}}$ being Clebsch-Gordan coefficients
($0 \le K \le 2S$).  These tensors form an orthonormal basis and have
the right properties under SU(2) transformations.  The crucial point
is that $\op{T}^{(S)}_{Kq}$ can be jotted down in terms of the $K$th
power of the Stokes operators.

The expansion coefficients
$\varrho_{Kq}^{(S)} = \Tr [ \op{\varrho}^{(S)} \, 
\op{T}_{Kq}^{(S) \, \dagger} ]$
are known as state multipoles.  The quantity
$\sum_{q} \mid \varrho_{Kq}^{(S)} \mid ^{2} $ gauges the state
overlapping with the $K$th multipole pattern.  For most states, only a
limited number of multipoles play a substantive role and the rest of
them have an exceedingly small contribution. Therefore, it seems more
convenient to look at the cumulative distribution~\cite{Hoz:2013om}
\begin{equation}
  \label{eq:cum}
  \mathcal{A}^{(S)}_{M} = \sum_{K= 1}^{M} \sum_{q=- K}^{K}
  \mid \varrho_{Kq}^{(S)} \mid ^{2}  \, ,
\end{equation}
which sums polarization information up to order $M$
($1 \le M \le 2S$). Note that the monopole $K=0$ is omitted, as it is
just a constant term. As with any cumulative distribution,
$ \mathcal{A}^{(S)}_{M} $ is a monotonically nondecreasing function of
the multipole order.

\emph{Maximal states.---} 
The distribution $\mathcal{A}^{(S)}_{M}$ can be regarded as a
nonlinear functional of the density matrix $\op{\varrho}^{^{(S)}}$. On
that account, one can try to ascertain the states that maximize
$\mathcal{A}^{(S)}_{M}$  for each order $M$.  We shall be
considering only pure states, which we expand as $|\Psi \rangle
=\sum_{m=-S}^{S} \Psi_{m} \, |S,m\rangle$, with coefficients $\Psi_{m}
= \langle S, m | \Psi \rangle$.  We easily get 
\begin{equation}
\label{eq:AMS}
\mathcal{A}_{M}^{(S)} =\sum_{K=1}^{M} \sum_{q=-K}^{K}
\frac{2K+1}{2S+1}
\left |  \sum_{m,m^{\prime }=-S}^{S} C_{Sm,Kq}^{Sm^{\prime} }
\Psi_{m^{\prime}} \Psi_{m}^{\ast } \right |^{2} \, .
\end{equation}
The details of the calculation are presented in the Supplemental
Material. We content ourselves with the final result: the maximum
value is
\begin{equation}
  \mathcal{A}_{M}^{(S)} =  \frac{2S}{2S +1} -
  \frac{[\Gamma (2S + 1)]^{2}}{\Gamma (2S-M) \Gamma (2S + M +2)} \, ,
\end{equation}
and this happens for the state $| S, \pm S\rangle$, irrespective of
$M$. Since $\mathcal{A}^{(S)}_{M}$ is invariant under polarization
transformations, all the displaced versions
$| \theta, \phi \rangle = (1 + | \alpha|^{2})^{-S} \exp(
\alpha\op{S}_{+} ) |S, -S \rangle$
[with $\op{S}_{\pm} = \op{S}_{x} \pm i \op{S}_{y}$ and the
stereographic projection $\alpha = \tan (\theta/2) e^{-i \phi}$] also
maximize $\mathcal{A}_{M}^{(S)}$. In other words, SU(2) coherent
states $| \theta, \phi \rangle$~\cite{Perelomov:1986ly} maximize
$\mathcal{A}^{(S)}_{M}$ for all orders $M$.

It will be useful in the following to exploit the Majorana
representation~\cite{Majorana:1932ul}, which maps every
$(2S+1)$-dimensional pure state $| \Psi \rangle$ into the polynomial
\begin{equation}
  \label{eq:MajPol}
  \Psi ( \alpha) = \sum_{m=-S}^{S} \sqrt{\frac{(2S)!}{(S-m)! (S+m)!}}
  \Psi_{m} \, \alpha^{S+m} \, .
\end{equation}
Up to a global unphysical factor, $| \Psi \rangle$ is determined by
the set $\{ \alpha_{i} \}$ of the $2S$ complex zeros of
$\Psi (\alpha)$, suitably completed by points at infinity if the
degree of $\Psi (\alpha) $ is less than $2S$.  A nice geometrical
representation of $| \Psi \rangle$ by $2S$ points on the unit sphere
(often called the constellation) is obtained by an inverse
stereographic map of
$\{ \alpha_{i}\} \mapsto \{ \theta_{i}, \phi_{i} \}$. For SU(2)
coherent states, the Majorana constellation collapses to a single
point. States with the same Majorana constellation, irrespective of
its relative orientation, share the same polarization properties.

The SU(2) $Q$-function, defined as
$Q(\theta, \phi) = \mid \langle \theta, \phi | \Psi \rangle \mid^{2}$,
is an alternative way to depict the state. Although $Q(\theta, \phi)$
can be expressed in terms of the Majorana polynomial [and so
$\{ \alpha_{i} \}$ are also the zeros of $Q (\theta, \phi)$],
sometimes the symmetry group of $|\Psi \rangle$ can be better
appreciated with this function, which can be very valuable.

\emph{Minimal states.---} 
Next, we concentrate on minimizing $\mathcal{A}_{M}^{(S)}$. Obviously,
the maximally mixed state
$\op{\varrho}^{(S )} = \textstyle{\frac{1}{2S+1}}
\op{\openone}_{2S+1}$
kills all the multipoles and so indeed causes \eqref{eq:cum} to vanish
for all $M$, being fully
unpolarized~\cite{Prakash:1971fr,Agarwal:1971zr}.  Nonetheless, we are
interested in pure $M$th-order unpolarized states.  The strategy we
adopt is thus very simple to state: starting from a set of unknown
normalized state amplitudes in Eq.~\eqref{eq:AMS}, which we write as
$\Psi_{m} = a_{m}+i b_{m}$ ($a_{m}, b_{m} \in \mathbb{R}$), we try to
get $\mathcal{A}_{M}^{(S)}=0$ for the highest possible $M$.  This
yields a system of polynomial equations of degree two for $a_{m}$ and
$b_{m}$, which we solve using Gr{\"o}bner bases implemented in the
computer algebra system {\sc magma}~\cite{Bosma:1997xp}.  In this way, we
get exact algebraic expressions and we can detect when there is no
feasible solution.

\begin{figure*}
  \centerline{\includegraphics[width=2.05\columnwidth]{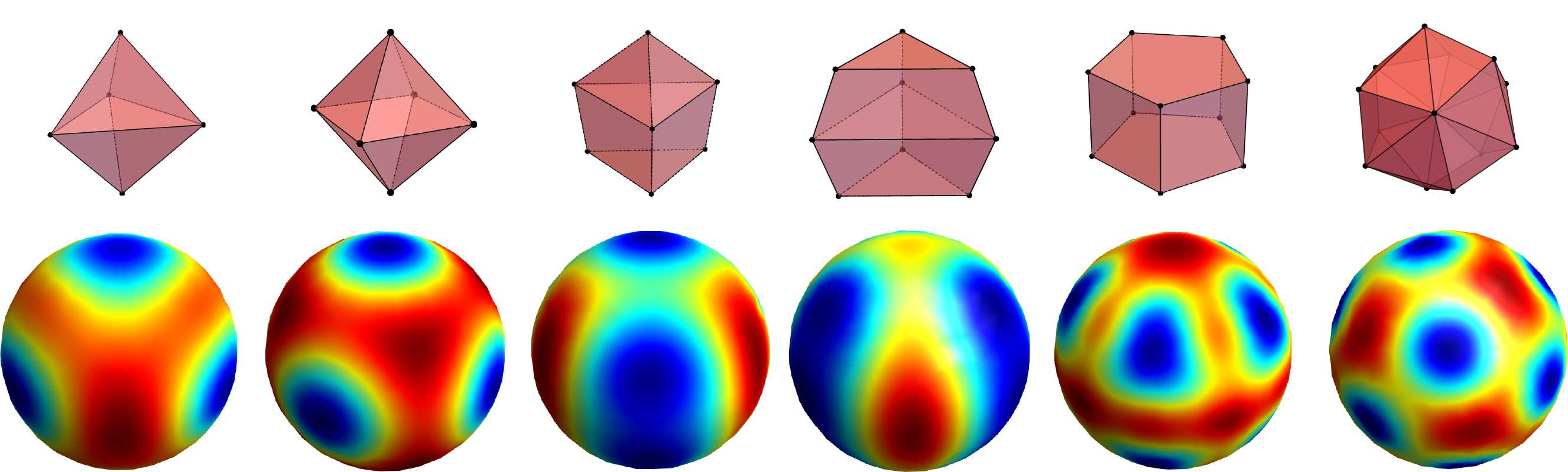}}
  \caption{(Color online) Density plots of the SU(2) $Q$ functions for
    the optimal states in Table I for the cases $S = 5/2,  3,
    7/2, 9/2,  5$, and $7$ (from left to right, blue
    indicates the zero values and red maximal ones). On top, we sketch
    the Majorana constellation for each of them.}
  \label{fig:Qfunc}
\end{figure*}

Table~\ref{table1} lists the resulting states (which, in some cases,
are not unique) for different selected values of $S$~\footnote{A more
  detailed list of minimal states can be found in
  http://polarization.markus-grassl.de}.  We also indicate the
associated Majorana constellations. For completeness, in
Fig.~\ref{fig:Qfunc} we also plot the constellations as well as the
$Q$-functions for some of these states.

Intuitively, one would expect that these constellations should have
the points as symmetrically placed on the unit sphere as possible.
This fits well with the notion of states of maximal Wehrl-Lieb
entropy~\cite{Baecklund:2014ng}. In more precise mathematical terms,
such points may be generated via optimization with respect to a
suitable criterion~\cite{Saff:1997gj}. Here, we explore the connection
with spherical $t$-designs~\cite{Delsarte:1977dn}, which are patterns
of $N$ points on a sphere such that every polynomial of degree at most
$t$ has the same average over the $N$ points as over the sphere. Thus,
the $N$ points mimic a flat distribution up to order $t$, which
obviously implies a fairly symmetric distribution.

For a given $S$, the maximal order of $M$ for which we can cancel out
$\mathcal{A}_{M}^{(S)}$ does not follow a clear pattern. The numerical
evidence suggests that $M_{\mathrm{max}}$ coincides
with $t_{\mathrm{max}}$ in the corresponding spherical design, but
further work  is needed to support this conjecture.

%%%%%%%%%%%%%%%%%%%%%%%%%%%%%%%%%%%%%%%%%%%%
\begin{table*}[t]
  \caption{States that kill $\mathcal{A}_{M}^{(S)}$ for the
    indicated values of $S$. In the second column, we indicate the
    order $M$, which we conjecture is the highest possible. We give 
    the nonzero state components $\Psi_{m}$ ($m=-S, \ldots, S$) and 
    the  Majorana  constellation. We include the associated spherical 
    $t$-design (with the maximal $t$ value) and the Queens of
    quantumness (with their unpolarization degree). ``same''
    ``similar'' and ``different'' always refer to the closest description column to the
    left.}
  \label{table1}
  \begin{ruledtabular}
    \begin{tabular}{cccccccc}
      $S$ & $M$ &  State & Constellation & Design & $t$ & Queens & $M$ \\
      \hline
 1 & 1 & $\Psi_{0} = 1$ & radial line & same & 1 & same & 1\\
 $\frac{3}{2}$ & 1 & $\Psi_{- \scriptsize{\frac{3}{2}}} = 
\Psi_{\scriptsize{\frac{3}{2}}} = \scriptsize{\frac{1}{\sqrt{2}}}$ & 
equatorial triangle  & same & 1 & same & 1\\
 2 & 2 & $\Psi_{-1} = \scriptsize{\frac{1}{\sqrt{3}}} , \quad 
\Psi_{2} = \scriptsize{\sqrt{\frac{2}{3}}} $ & 
tetrahedron  & same & 2 & same & 2 \\
$\frac{5}{2}$ & 1 & $\Psi_{- \scriptsize{\frac{5}{2}}} = 
\Psi_{\scriptsize{\frac{5}{2}}} = \scriptsize{\frac{1}{\sqrt{2}}}$ & 
 equatorial triangle  + poles  & same & 1 & same & 1 \\
 3 & 3 & $\Psi_{-2} = \Psi_{2} = \scriptsize{\frac{1}{\sqrt{2}}}$  &
 octahedron  & same & 3 & same & 3\\
 $\frac{7}{2}$ & 2 & $\Psi_{- \scriptsize{\frac{5}{2}}}=
 \Psi_{\scriptsize{\frac{1}{2}}} =   \scriptsize{\sqrt{\frac{7}{18}}},
\quad 
\Psi_{ \scriptsize{\frac{7}{2}}} = \scriptsize{\sqrt{\frac{2}{9}}} $ &  
two triangles + pole   & similar & 2 & equatorial pentagon + poles  & 1 \\
4 & 3 & $\Psi_{-4} =\Psi_{4} = \scriptsize{\sqrt{\frac{5}{24}}}, \quad
  \Psi_{0} = \scriptsize{\sqrt{\frac{7}{12}}} $ & cube  & same & 3 & 
 see~\cite{Giraud:2010db}  & 1 \\
$\frac{9}{2} $ & 2 & $\Psi_{- \scriptsize{\frac{9}{2}}} =
\Psi_{\scriptsize{\frac{9}{2}}} =  \scriptsize{\frac{1}{\sqrt{6}}} ,  \quad
  \Psi_{- \scriptsize{\frac{3}{2}}} = \Psi_{\scriptsize{\frac{3}{2}}}
 = \scriptsize{\frac{1}{\sqrt{3}}} $ 
& three triangles  & similar & 2 & similar & 1 \\
5 & 3 & $\Psi_{-5}= \Psi_{5} =  \scriptsize{\frac{1}{\sqrt{3}}} ,
\quad 
\Psi_{0} =  \scriptsize{\frac{1}{\sqrt{5}}}$ & 
pentagonal prism   & similar & 3 & two staggered squares + poles  & 1\\
$\frac{11}{2} $ & 3 & $\Psi_{- \scriptsize{\frac{11}{2}}} =
\Psi_{\scriptsize{\frac{11}{2}}}=  \scriptsize{\frac{\sqrt{17}}{12}}  ,
\quad \Psi_{- \scriptsize{\frac{5}{2}}} = 
\Psi_{\scriptsize{\frac{5}{2}}} = i \scriptsize{\frac{\sqrt{55}}{12}}$  & 
pentagon + two triangles   & similar & 3 & similar & 1 \\ 
6 & 5 &$\Psi_{-5}= -\Psi_{5}= \scriptsize{\frac{\sqrt{7}}{5}}, 
 \quad \Psi_{0} =- \scriptsize{\frac{\sqrt{11}}{5}} $ & 
icosahedron   & same & 5 & same & 5\\
7 & 4 & $\Psi_{-6}= \Psi_{6}=  \scriptsize{\sqrt{\frac{854}{3645}}},  \quad
\Psi_{-3} =  \Psi_{3}=  \scriptsize{\sqrt{\frac{637}{13420}} + i 
\sqrt{\frac{512603}{9783180}}} $ & 
three squares + poles & different & 4 & --- & -- \\
&  &  $\Psi_{0} = \scriptsize{\sqrt{\frac{12561757}{163053000}}} -
     \scriptsize{i \sqrt{\frac{512603}{2013000}}}$ & & & & \\
10 & 5 & $\Psi_{-10}= \Psi_{10} = \scriptsize{\sqrt{\frac{187}{1875}}} , \quad 
\Psi_{-5}=  -\Psi_{5}=\scriptsize{\sqrt{\frac{209}{625}}} ,  \quad 
\Psi_{0} =\scriptsize{\sqrt{\frac{247}{1875}}}$ & 
deformed dodecahedron & similar & 5& --- & --
    \end{tabular}
  \end{ruledtabular}
\end{table*}
%%%%%%%%%%%%%%%%%%%%%%%%%%%%%%%%%%%%%%%%%%%%%

The simplest nontrivial example are two-photon states, $S=1$.  We find
only first-order unpolarized states: these are biphotons
generated in spontaneous parametric down-conversion, which
were the first known to have hidden
polarization~\cite{Klyshko:1997yq}.

With three photons, $S=3/2$, we have again only first-order
unpolarized states: the constellation is an equilateral triangle
inscribed in a great circle, which can be taken as the equator. This
coincides with the three-point spherical 1-design.

For $S=2$, the Majorana constellation is a regular tetrahedron: it is
the least-excited second-order unpolarized state. It is not
surprising that the tetrahedron is the $2$-design with the lowest
number of points.

The case $S= 5/2$ does not admit a high degree of spherical symmetry:
only first-order unpolarized states exist. There are neither five-photon
$M=2$ unpolarized states~\cite{Kolenderski:2008mo,Crann:2010qd} nor
five-point $2$-designs~\cite{Mimura:1990xg}.

When increasing the number of photons to six, $S=3$, another Platonic
solid appears: the regular octahedron. Now, we have the least-excited
third-order unpolarized states, which, in addition, take on the maximum sum
of the Stokes variances.  

For $S= 7/2$, an $M=2$ constellation consists of the north pole, an
equilateral triangle inscribed at the $z = 0.2424$
plane, and another equilateral triangle, with the same orientation
(e.g., one vertex on the $x$-axis) at the $z =  - 0.5816$ plane.  The
spherical $t$-design has a larger separation between the
triangles; but the corresponding Stokes vector does not
vanish, so the $t$-design does not coincide with any unpolarized state.

%%%%%%%%%%%%%%%%%%%%%%%%%%%%%%%%%%%%%%%%%%%%%%%%%
%%%%%%%%%%%%%%%%%%%%%%%%%%%%%%%%%%%%%%%%%%%%%%%%%%

The next Platonic solid, the cube, appears when $S=4$. The state is
third-order unpolarized and its Majorana constellation coincides with
the eight-point spherical $3$-design, which is the tightest for this
number of points.

A nine-photon second-order unpolarized state, $S=9/2$, is generated by
three equilateral triangles with the same orientation inscribed in the
equator and in two symmetric rings.  The highest nine-point spherical
$t$-design has $t = 2$ and a similar, but not identical, configuration
because the two smaller triangles are displaced by a larger distance
from the equator than the previous constellation. As a consequence,
the nine-point spherical $2$-design is only first-order unpolarized.

The Majorana constellation for a maximally unpolarized 10-photon
state ($S=5$) is similar to the matching spherical $t$-design and
consists of two identical regular pentagons inscribed in rings
symmetrically displaced from the equator. The maximally unpolarized
state has the two pentagons a bit closer to the equator than the
spherical $3$-design (that has $M=1$).

For larger photon numbers, the computational complexity of finding
optimal designs becomes a real hurdle.  The $t$-designs have been
investigated in the range 2--100 and numerical evidence suggests that
the optimal designs (in some instances, they are not unique) have been
found~\footnote{For a complete account see
  http://neilsloane.com/sphdesigns/}. However, for some dimensions,
e.g., 12 ($S=6$) and 20 ($S=10$), one would na\"{i}vely guess that the
optimal designs fit with the icosahedron and the dodecahedron. For
$S=6$ this turns out to be a correct guess, the corresponding state is
unpolarized to the same order as the spherical $5$-design formed by
the icosahedron.  For $S=10$ this intuition fails: the optimal
$t$-design is indeed a dodecahedron, but this Majorana constellation
is third-order unpolarized, whereas this is a spherical $5$-design. If
the dodecahedron is stretched (i.e., the four pentagonal rings that
define its vertices are displaced against the pole), one can find a
20-photon fifth-order unpolarized state.

To check the correspondence between unpolarized states and optimal
$t$-designs we look at dimension 14, which is the smallest number of
points for which a spherical $4$-design, but not a $5$-design,
exists. This consists of four equilateral triangles that are pairwise
similar in size, displaced from the equator by the same distance, and
rotated an angle $\pm \alpha$ or $\pm \beta$ around their surface
normal, plus the two poles. The $t$-design state is only first-order
unpolarized, but if the spacing and triangle orientation is optimized,
the design can be made third-order unpolarized. There is indeed a
14-photon state that is fourth-order unpolarized: its Majorana
constellation is made of three quadrangles and the poles, but this is
only a $1$-design.

To round up, it is worth commenting on the connections that our theory
shares with two recently introduced notions: anticoherent
states~\cite{Zimba:2006fk} and queens of
quantumness~\cite{Giraud:2010db}. For completeness, in
Table~\ref{table1} we have also listed the configurations and the
degree of unpolarization for these queens.  Anticoherent states are in
a sense ``the opposite'' of SU(2) coherent states: while the latter
correspond as nearly as possible to a classical spin vector pointing
in a given direction, the former ``point nowhere'', i.e., the average
Stokes vector vanishes and the fluctuations up to order $M$ are
isotropic.  The queens of quantumness are the most distant states (in
the sense of a Hilbert-Schmidt distance) to the classical ones (states
than can be written as a convex sum of projectors onto coherent
states). In particular low-dimensional cases, these two instances
coincide with our optimal states. However, we stress that our theory
is built from first principles, starting from magnitudes that are
routinely determined in the lab. Besides, we have an algebraic
criterion, namely, the vanishing of the cumulative multipole
distribution, that can be handled in a clear and compact manner.

When we interpret our $(2S+1)$-dimensional subspace as the symmetric
subspace of a system of $S$ qubits, the kings appear also closely
linked to other intriguing problems, such as maximally entangled
symmetric states~\cite{Aulbach:2010jw,Giraud:2015oj} and $k$-maximally
mixed states~\cite{Arnaud:2013hm,Goyeneche:2014so}.

\emph{Applications.---}
The main goal of quantum metrology is to measure a physical magnitude
with surprising precision by exploiting quantum resources. In
particular, tailoring polarization states to better detect SU(2)
rotations is quite a relevant problem with direct applications to 
magnetometry, polarimetry, and metrology, in
general~\cite{Rozema:2014fk}.

In this respect, $N00N$ states [defined as
$ |N00N \rangle = (|S ,S \rangle - | S,-S \rangle)/\sqrt{2}$] are
known to be maximally sensitive to small phase shifts (i.e., to
small rotations about the $S_{z}$-axis) for a fixed excitation
$S$~\cite{Bollinger:1996am}. This can be easily understood by looking at
their Majorana constellation, which consists in just $2S$
equidistantly placed points around the Poincar\'{e} sphere
equator. Since a rotation around the $S_{z}$ axis is described by the
unitary operator $\hat{U}(\vartheta)= \exp(- i \vartheta
\op{S}_{z}/2)$,  the states $| \mathrm{N00N} \rangle$ and
$\hat{U}(\vartheta) | \mathrm{N00N} \rangle$ are orthogonal for
$\pi/(2S)$. However, to make optimal use of a $N00N$ state it is
essential to know the rotation axis so as to ensure that the state is
aligned with the axis to achieve its best sensitivity: the rotation
resolution is thus highly directional for a $N00N$ state.

This is precisely the advantage of maximally unpolarized states:
having a high degree of spherical symmetry, they resolve rotations
around any axis approximately equally well. This has been confirmed
for the Platonic solids~\cite{Kolenderski:2008mo}: Platonic
states saturate the optimal average sensitivity to rotations about any
axis; $N00N$ states outperform these states about one specific
axis~\cite{Rozema:2014bh}. Indeed, for the Platonic solids, rotations
around all the facets normal axes map the Majorana constellation onto
itself for rotations of $2 \pi/3$ (tetrahedron, octahedron and
icosahedron), $\pi/2$ (cube), or $2 \pi/5$ (dodecahedron). It is clear
that for other constellations and other rotation axes the Majorana
constellation will only become approximately identical, but the
statement is more likely to hold true.

In a different vein, we draw  attention to the structural
similarity between the kings of quantumness and quantum error
correcting codes: in both cases, low-order terms in the expansion of
the density matrices are required to vanish.

As a final but relevant remark, we stress that all the basic tools
needed for our treatment (Schwinger representation, multipole
expansion and constellations) have been extended in a direct way to
other symmetries, such as SU(3)~\cite{Banyai:1966dq} or
Heisenberg-Weyl~\cite{Ivan:2012hb}.  Therefore, the notion of kings of
quantumness can be easily developed for other systems. Work along
these lines is already in progress in our group.

\emph{Concluding remarks.---}
In short, we have consistently  reaped the benefits of the 
cumulative distribution of polarization multipoles, which is a
sensible and experimentally realizable quantity. We have proven that
SU(2) coherent states maximize that quantity to all orders: in this
way, they manifest their classical virtues. Their opposite
counterparts, minimizing that quantity, are certainly the kings of
quantumness 

Apart from their indisputable geometrical beauty, there surely
is plenty of room for the application of these states, whose
generation has started to be seriously considered in several groups.

The authors acknowledge interesting discussions with Prof. Daniel
Braun and Olivia di Matteo. Financial support from the Swedish
Research Council (VR) through its Linnaeus Center of Excellence ADOPT
and Contract No. 621-2011-4575, the CONACyT (Grant 106525), the
European Union FP7 (Grant Q-ESSENCE), and the Program UCM-Banco
Santander (Grant GR3/14) is gratefully acknowledged. GB thanks the MPL
for hosting him and the Wenner-Gren Foundation for economic support.

\appendix

\emph{Appendix: Optimal states.---}
We have to maximize the cumulative multipole distribution \eqref{eq:cum}
for a pure state $| \Psi \rangle =\sum_{m=-S}^{S} \Psi_{m}
\,|S,m\rangle$, which takes the form \eqref{eq:AMS}. 
 If we use integral representation for the product of two
 Clebsch-Gordan coefficients~\cite{Varshalovich:1988ct}, we get 
\begin{gather}
\mathcal{A}_{K}^{(S)} = \sum_{m,m^{\prime }=-S}^{S} 
\sum_{n,n^{\prime} =-S}^{S}  \frac{2S+1}{8\pi ^{2}} 
\sum_{K=0}^{M}\sum_{q=-K}^{K}\frac{2K+1}{2S+1} 
\Psi_{m^{\prime }} \Psi_{m}^{\ast } \Psi_{n} \Psi_{n^{\prime }}^{\ast }
\nonumber \\
\times \int
dR \, D_{mn}^{S} (R ) \,  
D_{m^{\prime }n^{\prime }}^{S \ast } ( R ) \, 
D_{qq}^{K} (R) \,,
\label{1}
\end{gather}
where $D_{mn}^{S}$ are the Wigner $D$-functions and $R$ refers to the
three Euler angles $(\alpha, \beta, \gamma)$ and the integration is on the
group manifold
\begin{equation}
\int dR \ f(R) \equiv
\int_{0}^{2 \pi} d\alpha \int_{0}^{\pi} d\beta \sin \beta
\int_{0}^{2\pi} d\gamma f(\alpha, \beta, \gamma) \ .
\end{equation}
Since
\begin{equation}
\sum_{q=-K}^{K} D_{qq}^{K} (R) =\chi _{K} (\omega ),
\end{equation}
where $\chi_{K} (\omega)$ is a SU(2) generalized character 
and $\cos (\omega/2) = \cos (\beta/2) \cos [(\alpha + \gamma)/2]$,
we rewrite $\mathcal{A}_{M}^{(S)}$ as
\begin{equation}
\mathcal{A}_{M}^{(S)} = \sum_{K=0}^{M} \frac{2K+1}{8\pi ^{2}}
\int dR \, \chi_{K}(\omega ) \mid \langle \Psi | \op{T}_{g}^{S} |\Psi \rangle \mid^{2}.
\end{equation}
and $\op{T}_{g}$ is the group action.
Then, we observe that the above is
\begin{gather}
\mathcal{A}_{M}^{(S)} = \Tr \Big [  |\Psi \rangle \langle \Psi |\otimes
  |\tilde{\Psi} \rangle  \langle \tilde{\Psi} |  \nonumber \\
\times \sum_{K=0}^{M}\frac{2K+1}{8\pi^{2}} 
\int dR \, \chi_{K}(\omega ) \op{T}_{g}^{S}\otimes
  \op{T}_{g}^{S} \Big] ,
\end{gather}
with  
\begin{equation}
|\tilde{\Psi} \rangle = \sum_{m=-S}^{S}  (-1)^{m} \, \Psi_{-m}^{\ast} \,
|S,m \rangle .  \label{psi1}
\end{equation}
 
Because the integral  
\begin{eqnarray}
\frac{1}{4\pi ^{2}}\int dR \, \chi _{K} (\omega ) \op{T}_{g}^{S}\otimes 
\op{T}_{g}^{S\dagger} =  c_{K}\Pi_{K},  \label{2} 
\end{eqnarray}
where $\Pi _{K}$ is the identity on the $(2K+1)$-dimensional irreducible SU(2)
subspace which appear in the tensor product of $\mathcal{H}_{S}\otimes
\mathcal{H}_{S}$ [i.e., $\Tr ( \Pi _{K} ) = 2K+1$], then 
\begin{equation}
\mathcal{A}_{M}^{(S)}=\sum_{K=1}^{M}\langle \tilde{\Psi} |
\langle \Psi |  \Pi_{K}| \Psi \rangle | \tilde{\Psi} \rangle .
\end{equation}
Such overlap is maximized (all coefficients are the same) whenever in every
subspace of dim $2K+1$ there is only one element from the decomposition $
|\Psi \rangle |\tilde{\Psi}\rangle $, which is consistent with (\ref{psi1}). The
only states that at decomposition on representations produce a single state
in each invariant subspace are the basis states $|S,m\rangle $, so
that   $|\tilde{\Psi} \rangle =(-1)^{m}  \, |S,-m\rangle $, then 
 \begin{equation}
\mathcal{A}_{M}^{(S)} = \sum_{K=1}^{M} 
\frac{2K+1}{2S+1} \left\vert C_{SS,K0}^{S-S}\right \vert^{2} \, .
\end{equation}
Since the maximum value of $C_{Sm_{,}S-m}^{K0}$
is $C_{SS,S-S}^{K0}$, the states $|S,\pm S\rangle $ maximize
$\mathcal{A}_{M}^{(S)}$, as heralded  before.

%\bibliography{Polarization}
%merlin.mbs apsrev4-1.bst 2010-07-25 4.21a (PWD, AO, DPC) hacked
%Control: key (0)
%Control: author (0) dotless jnrlst
%Control: editor formatted (1) identically to author
%Control: production of article title (0) allowed
%Control: page (1) range
%Control: year (0) verbatim
%Control: production of eprint (0) enabled
%

\end{document}